\documentclass[aps,prb,twocolumn,showpacs]{revtex4-1}

\usepackage{graphicx}

\begin{document}

%Title of paper
\title{\emph{Ab initio} calculations of phonon spectra in \emph{A}TiO$_3$
perovskite crystals\\
(\emph{A} = Ca, Sr, Ba, Ra, Cd, Zn, Mg, Ge, Sn, Pb) }

\author{Alexander I. Lebedev}
\email[]{swan@scon155.phys.msu.ru}
\affiliation{Physics Department, Moscow State University,
119991 Moscow, Russia}

\date{\today}

\begin{abstract}
The phonon spectra of calcium, strontium, barium, radium, cadmium, zinc,
magnesium, germanium, tin, and lead titanates with the perovskite structure
are calculated from first principles within the density functional theory.
By analyzing the unstable modes in the phonon spectra, the possible lattice
distortions are determined and the energies of the corresponding phases are
calculated. From analyzing the phonon spectra, force constants, and eigenvectors
of TO phonons, a conclusion is drawn on the origin of the ferroelectricity
in considered crystals. It is shown that the main factors determining the
possible off-centering of atoms in the \emph{A}~position are the geometric size
and electronic configuration of these atoms.

\texttt{DOI: 10.1134/S1063783409020279}

\end{abstract}

% insert suggested PACS numbers in braces on next line
\pacs{61.50.Ah 63.20.Dj 71.15.Mb 77.84.Dy}

\maketitle

\section{Introduction}

Crystals of the perovskite family are well-known materials undergoing various
structural distortions with decreasing temperature. When the character of these
distortions is ferroelectric, a number of physical properties of these crystals
(dielectric constant, piezoelectric coefficients, etc.) becomes anomalously
large. For this reason, these materials have found wide application in modern
electronics.

The problem of further optimization of the ferroelectric properties requires
a deeper understanding of the microscopic mechanisms responsible for the
appearance of ferroelectricity and the ferroelectric properties. In solving
this problem, very useful information can be obtained from the \emph{ab initio}
calculations which have already made a
significant contribution to the understanding of the ferroelectric phenomena
in the perovskite crystals.~\cite{PhysRevB.42.6416,Nature.358.136,
PhysRevLett.72.3618,PhysRevB.49.5828,PhysRevLett.74.4067,
Ferroelectrics.194.109,PhysRevB.55.6161,Ferroelectrics.206.205,PhysRevB.60.836,
ApplPhysLett.81.3443,PhysSolidState.44.1135}

In discussing the properties of ferroelectrics, it is very important to
understand whether these properties result from a collective displacement of
atoms in a lattice (displacive phase transition) or they are due to specific
features of certain constituent atoms in a crystal (order--disorder phase
transition). This problem arises, in particular, in discussing the nature
of the phase transitions that occur in incipient ferroelectrics doped with
certain impurities.~\cite{Lemanov2000}

Earlier studies of titanates with the perovskite structure have dealt mainly
with four compounds (CaTiO$_3$,~\cite{PhysRevLett.72.3618,PhysRevB.49.5828,
PhysRevB.62.3735,JChemPhys.114.2395}
SrTiO$_3$,~\cite{PhysRevLett.72.3618,PhysRevB.49.5828,Ferroelectrics.194.109}
BaTiO$_3$,~\cite{PhysRevB.42.6416,Nature.358.136,PhysRevLett.72.3618,
PhysRevB.49.5828,Ferroelectrics.206.205,PhysRevB.60.836}
and PbTiO$_3$~\cite{Nature.358.136,PhysRevLett.72.3618,PhysRevB.49.5828,
PhysRevB.55.6161,PhysRevB.60.836}) and their solid solutions. However, a
comparison of the results obtained in these studies is hampered by different
calculation techniques used in these investigations and different methods used
to construct atomic (pseudo)potentials.

The aim of this work is to carry out first-principles calculations of
the phonon spectra of ten \emph{A}TiO$_3$ crystals with the perovskite structure
and determine the structure of the most stable phases. In order to
test our approach, we first apply it for calculating the properties of the four
above-mentioned systems. Then we predict the properties of poorly investigated
or hypothetical perovskite compounds RaTiO$_3$, CdTiO$_3$, MgTiO$_3$, ZnTiO$_3$,
SnTiO$_3$, and GeTiO$_3$.%
    \footnote{The properties of one more member of this family of titanates,
    HgTiO$_3$, have been considered in Phys. Solid State \textbf{54}, 1663
    (2012); \texttt{DOI: 10.1134/S1063783412080185}.}
From comparison of the results obtained in a unified way
for a large number of related materials, conclusions are made about the relation
of the structural distortions in \emph{A}TiO$_3$ crystals to the size and
electronic structure of the \emph{A} atom. From the analysis of the on-site
force-constant matrix and the TO-phonon eigenvectors, we draw conclusions
about the origin of the ferroelectricity in these materials and find conditions
under which the ferroelectric phenomena can be associated with off-center atoms.

\section{Calculation technique}

The calculations were performed using the ABINIT software~\cite{Abinit} based
on the density functional theory, pseudopotentials, and the plane-wave expansion
of the wave functions. The exchange-correlation interaction was described
in the local density approximation (LDA).~\cite{PhysRevB.23.5048}
The used pseudopotentials were optimized separable nonlocal
pseudopotentials,~\cite{PhysRevB.41.1227} which were constructed using the OPIUM
code and to which the local potential correction~\cite{PhysRevB.59.12471}
was added in order to improve their transferability. For elements with atomic
numbers $Z < 46$, the construction was performed in a non-relativistic way; for
other elements, a scalar-relativistic approximation was used. Table~\ref{Table1}
lists the parameters used for constructing pseudopotentials. The local potential
was the $s$ potential except for the oxygen atom, for which the local $d$
potential was used. The parameters of pseudopotentials were finely adjusted by
comparing the calculated and experimental values of the lattice parameters for
a number of oxides and sulfides of the elements.

The lattice parameters and the equilibrium atomic positions in the unit cell
were obtained by minimizing the Hellmann--Feynman forces acting on the atoms
($<$10$^{-5}$ Ha/Bohr), with the total energy being calculated self-consistently
with an accuracy of better than 10$^{-10}$~Ha.%
    \footnote{In this paper, the energy is measured in Hartrees
    (1~Ha = 27.2113845~eV) everywhere except Tables~\ref{Table1} and
    \ref{Table3}.}
In the calculations, particular attention was paid to the convergence of the
results with respect to the kinetic energy cut-off for plane waves and the
density of the $k$-point mesh used in the integration over the Brillouin zone.
For all the calculated properties presented below, the convergence was attained
at an energy cut-off of 30~Ha and the 8$\times$8$\times$8 $k$-point mesh
constructed according to Ref.~\onlinecite{PhysRevB.13.5188}.

The Born effective charges $Z^*$, optical dielectric constant $\epsilon_\infty$,
elastic moduli $C_{ij}$, bulk modulus $B$, force-constant matrix $\Phi_{ij}$,
and phonon spectra were calculated using the density-functional perturbation
theory.~\cite{PhysRevB.43.7231,
PhysRevB.55.10337,PhysRevB.55.10355,PhysRevB.71.035117} The phonon frequencies
were calculated exactly at five points of the Brillouin zone ($\Gamma$, $X$, $M$,
$R$, and the $\Lambda$ point located halfway between the $\Gamma$ and $R$ points),
and then the phonon spectrum was computed over the entire Brillouin zone using the
interpolation technique.~\cite{PhysRevB.43.7231,PhysRevB.50.13035}

\begin{table*}
\caption{\label{Table1}Electronic configurations of atoms and parameters used
to construct pseudopotentials: $r_s$, $r_p$, and $r_d$ are the core radii of
the pseudopotentials for the $s$, $p$, and $d$ projections; $q_s$, $q_p$, and
$q_d$ are the cut-off wave vectors used to optimize pseudopotentials;
and $r_{\rm min}$, $r_{\rm max}$, and $V_{\rm loc}$ are the limits and the
depth of the correcting local potential (parameter values are in atomic units,
and energy is given in Ry).}
\begin{ruledtabular}
\begin{tabular}{ccccccccccc}
Atom & Configuration & $r_s$ & $r_p$ & $r_d$ & $q_s$ & $q_p$ & $q_d$ & $r_{\rm min}$ & $r_{\rm max}$ & $V_{\rm loc}$ \\
\hline
Ca & $3s^23p^63d^04s^0$ & 1.46 & 1.68 & 1.82 & 7.07 & 7.07 & 7.27 & 0.01 & 1.40 & 1.6 \\
Sr & $4s^24p^64d^05s^0$ & 1.68 & 1.74 & 1.68 & 7.07 & 7.07 & 7.07 & 0.01 & 1.52 & 1.5 \\
Ba & $5s^25p^65d^06s^0$ & 1.85 & 1.78 & 1.83 & 7.07 & 7.07 & 7.07 & 0.01 & 1.68 & 1.95 \\
Ra & $6s^26p^67s^06d^07p^0$ & 1.84 & 1.73 & 1.98 & 7.8 & 7.8 & 7.8 & 0.01 & 1.68 & $-$1.3 \\
Mg & $2s^22p^63s^03p^0$ & 1.50 & 1.88 & ---  & 6.7  & 8.1  & ---  & 0.01 & 1.0  & $-$0.84 \\
Zn & $3d^{10}4s^04p^0$  & 1.82 & 1.82 & 2.00 & 7.07 & 7.07 & 7.47 & 0.01 & 1.60 & 2.5 \\
Cd & $4d^{10}5s^05p^0$  & 2.04 & 2.18 & 2.10 & 7.07 & 7.07 & 7.07 & 0    & 1.88 & $-$1.6 \\
Ge & $3d^{10}4s^{1.5}4p^{0.5}$ & 1.68 & 1.68 & 1.96 & 7.07 & 6.0 & 7.77 & 0.01 & 1.58 & 0.48 \\
Sn & $4d^{10}5s^25p^0$  & 2.14 & 2.08 & 2.18 & 7.07 & 7.07 & 7.07 & 0.01 & 1.90 & 0.64 \\
Pb & $5d^{10}6s^26p^0$  & 1.72 & 1.98 & 1.80 & 6.05 & 5.52 & 7.17 & 0.1  & 1.43 & 1.6 \\
Ti & $3s^23p^63d^04s^0$ & 1.48 & 1.72 & 1.84 & 7.07 & 7.07 & 7.07 & 0.01 & 1.41 & 2.65 \\
O  & $2s^22p^43d^0$     & 1.40 & 1.55 & 1.40 & 7.07 & 7.57 & 7.07 & ---  & ---  & --- \\
\end{tabular}
\end{ruledtabular}
\end{table*}

\section{Testing of the calculation technique}

The correctness of the described approach was tested by comparing the
calculated lattice parameters, spontaneous polarizations, and phonon spectra
with available experimental data and calculations performed by other authors
for well-studied CaTiO$_3$, SrTiO$_3$, BaTiO$_3$, and PbTiO$_3$ compounds.

\begin{table*}
\caption{\label{Table2}Comparison of calculated and experimental lattice
parameters of different phases of \emph{A}TiO$_3$ compounds (experimental data
are obtained at 300~K, unless otherwise specified).}
\begin{ruledtabular}
\begin{tabular}{cccc}
Compound  & Space group & Source & Lattice parameters \\
\hline
CaTiO$_3$ & $Pbnm$      & This work  & $a = 5.3108$, $b = 5.4459$, $c = 7.5718$~{\AA} \\
          &             & Exp.~\cite{LB} & $a = 5.3670$, $b = 5.4439$, $c = 7.6438$~{\AA} \\
SrTiO$_3$ & $Pm3m$      & This work  & $a = 3.8898$~{\AA} \\
          &             & Exp.~\cite{LB} & $a = 3.905$~{\AA} \\
          & $I4/mcm$    & This work  & $a = b = 5.4680$, $c = 7.8338$~{\AA} \\
          &             & Exp.~\cite{LB} & $a = b = 5.510$, $c = 7.798$~{\AA} (20~K) \\
BaTiO$_3$ & $Pm3m$      & This work  & $a = 3.9721$~{\AA} \\
          &             & Exp.~\cite{LB} & $a = 3.996$~{\AA} (393~K) \\
          & $P4mm$      & This work  & $a = 3.9650$, $c = 4.0070$~{\AA}, $c/a = 1.0106$ \\
          &             & Exp.~\cite{LB} & $a = 3.9920$, $c = 4.0361$~{\AA} (293~K) \\
          & $Amm2$      & This work  & $a = 3.9620$, $b = 5.6384$, $c = 5.6484$~{\AA} \\
          &             & Exp.~\cite{LB} & $a = 3.990$, $b = 5.669$, $c = 5.682$~{\AA} (263~K) \\
          & $R3m$       & This work  & $a = 3.9817$~{\AA}, $\alpha = 89.933^\circ$ \\
          &             & Exp.~\cite{LB} & $a = 4.001$~{\AA}, $\alpha = 89.85^\circ$ (105~K) \\
PbTiO$_3$ & $P4mm$      & This work  & $a = 3.8858$, $c = 4.1151$~{\AA}, $c/a = 1.0590$ \\
          &             & Exp.~\cite{LB} & $a = 3.904$, $c = 4.152$~{\AA}  \\
\end{tabular}
\end{ruledtabular}
\end{table*}

The lattice parameters corresponding to a minimum total energy of the crystals
are given in Table~\ref{Table2}. The obtained values agree well with the
experimental data~\cite{LB} if one takes into account that the LDA usually
slightly underestimates the lattice parameter. An analysis of the relative
energies (per formula unit) of low-symmetry phases (Table~\ref{Table3})
shows that for barium titanate, the most energetically favorable phase is the
rhombohedral one and for lead titanate, it is the tetragonal phase. The
calculated values of the $c/a$ ratio for tetragonal BaTiO$_3$ and PbTiO$_3$
are close to the experimental values (Table~\ref{Table2}). For CaTiO$_3$, the
most energetically favorable phase is the orthorhombic $Pbnm$ phase and for
SrTiO$_3$, it is the tetragonal $I4/mcm$ phase. The values of spontaneous
polarization calculated by the Berry phase method~\cite{PhysRevB.47.1651}
are 0.26, 0.31, and 0.89~C/m$^2$ for tetragonal BaTiO$_3$, rhombohedral
BaTiO$_3$, and tetragonal PbTiO$_3$,
respectively. These values are close to the experimental data (0.26, 0.33,
0.75~C/m$^2$, Ref.~\onlinecite{LB}).

\begin{table*}
\caption{\label{Table3}Relative energies of different low-symmetry phases of
\emph{A}TiO$_3$ compounds (energies of the most stable phases are in boldface).}
\begin{ruledtabular}
\begin{tabular}{cccccccccc}
Compound  & Unstable & Space & Energy & Unstable & Space & Energy & Unstable & Space & Energy \\
          & mode     & group & (meV)  & mode     & group & (meV)  & mode     & group & (meV) \\
\hline
MgTiO$_3$ & $\Gamma_{25}$ & $P{\bar 4}m2$ & $-$125$^a$ &
            $M_5^\prime$  & $Cmmm$   & $-$417$^a$ &
            $\Gamma_{15}$, $\Gamma_{25}$ & $Amm2$ & $-$1380$^a$ \\
          & $X_3$         & $P4_2/mmc$ & $-$147$^a$ &
            $X_5^\prime$  & $Pmma$   & $-$500 &
            $R_{25}+M_3$  & $Cmcm$   & $-$1658$^a$ \\
          & $R_{15}$      & $I4/mmm$ & $-$228$^a$ &
            $\Gamma_{25}$ & $R32$    & $-$686$^a$ &
            $R_{25}$      & $R{\bar 3}c$ & $-$1727 \\
          & $R_{15}$      & $R{\bar 3}m$ & $-$289$^a$ &
            $\Gamma_{15}$ & $R3m$    & $-$695 &
            $R_{15}$, $R_{25}$ & $Imma$   & $-$1764$^a$ \\
          & $X_5^\prime$  & $Cmcm$   & $-$304 &
            $\Gamma_{15}$ & $P4mm$   & $-$1028 &
            $R_{25}+M_3$  & $Pbnm$   & \textbf{$-$1992} \\
          & $M_5^\prime$  & $Pmma$   & $-$344$^a$ &
            $M_3$         & $P4/mbm$ & $-$1107 \\
          & $M_2^\prime$  & $P4/nmm$ & $-$417$^a$ &
            $R_{25}$      & $I4/mcm$ & $-$1111 \\
\hline
CaTiO$_3$ & $X_5$         & $Cmcm$   & $-$0.0$^a$ &
            $M_5^\prime$  & $Cmmm$   & $-$6.7 &
            $R_{25}$      & $I4/mcm$ & $-$365 \\
          & $X_5$         & $Pmma$   & $-$0.0$^a$ &
            $\Gamma_{15}$ & $R3m$    & $-$73.7 &
            $R_{25}$      & $R{\bar 3}c$ & $-$385 \\
          & $X_5^\prime$  & $Cmcm$   & $-$0.6 &
            $\Gamma_{15}$ & $Amm2$   & $-$85.4$^a$ &
            $R_{25}+M_3$  & $Cmcm$   & $-$404$^a$ \\
          & $X_5^\prime$  & $Pmma$   & $-$0.9 &
            $\Gamma_{15}$ & $P4mm$   & $-$123 &
            $R_{25}$      & $Imma$   & $-$412$^a$ \\
          & $M_5^\prime$  & $Pmma$   & $-$5.0 &
            $M_3$         & $P4/mbm$ & $-$321 &
            $R_{25}+M_3$  & $Pbnm$   & \textbf{$-$497} \\
\hline
SrTiO$_3$ & $\Gamma_{15}$ & $P4mm$   & $-$0.71 &
            $M_3$         & $P4/mbm$ & $-$9.45 &
            $R_{25}$      & $I4/mcm$ & \textbf{$-$30.9} \\
          & $\Gamma_{15}$ & $Amm2$   & $-$0.75$^a$ &
            $R_{25}$      & $R{\bar 3}c$ & $-$27.5 \\
          & $\Gamma_{15}$ & $R3m$    & $-$0.75 &
            $R_{25}$      & $Imma$$^b$ & $-$28.9 \\
\hline
BaTiO$_3$ & $M_3^\prime$  & $P4/nmm$ & $-$0.31 &
	    $X_5$         & $Cmcm$   & $-$1.45 &
	    $\Gamma_{15}$ & $Amm2$   & $-$7.4 \\
          & $X_5$         & $Pmma$   & $-$1.22 &
            $\Gamma_{15}$ & $P4mm$   & $-$5.6 &
	    $\Gamma_{15}$ & $R3m$    & \textbf{$-$8.1} \\
\hline
RaTiO$_3$ & $M_3^\prime$  & $P4/nmm$ & $-$11.1 &
            $X_5$         & $Cmcm$   & $-$16.9 &
	    $\Gamma_{15}$ & $Amm2$   & $-$28.5 \\
          & $X_5$         & $Pmma$   & $-$14.2 &
            $\Gamma_{15}$ & $P4mm$   & $-$21.8 &
	    $\Gamma_{15}$ & $R3m$    & \textbf{$-$29.7} \\
\hline
CdTiO$_3$ & $R_{15}$      & $I4/mmm$ & $-$24$^a$ &
	    $X_5$         & $Pmma$   & $-$160 &
            $R_{25}$      & $I4/mcm$ & $-$912 \\
          & $R_{15}$      & $R{\bar 3}m$ & $-$30$^a$ &
	    $\Gamma_{15}$ & $R3m$    & $-$245 &
            $M_3$         & $P4/mbm$ & $-$920 \\
          & $X_3$         & $P4_2/mmc$ & $-$45 &
	    $X_5$         & $Cmcm$   & $-$282 &
            $R_{25}+M_3$  & $Cmcm$   & $-$1157$^a$ \\
          & $X_5^\prime$  & $Cmcm$   & $-$66$^a$ &
            $\Gamma_{15}$ & $P4mm$   & $-$340 &
            $R_{25}$      & $R{\bar 3}c$ & $-$1197 \\
          & $X_5^\prime$  & $Pmma$   & $-$104$^a$ &
            $\Gamma_{15}$, $\Gamma_{25}$ & $Amm2$ & $-$412 &
            $R_{15}$, $R_{25}$ & $Imma$   & $-$1202$^a$ \\
          & $\Gamma_{25}$ & $P{\bar 4}m2$ & $-$134 &
            $\Gamma_{25}$ & $R32$    & $-$486 &
            $R_{25}+M_3$  & $Pbnm$   & \textbf{$-$1283} \\
\hline
ZnTiO$_3$ & $X_3$         & $P4_2/mmc$ & $-$171$^a$ &
            $M_5^\prime$  & $Cmmm$     & $-$564$^a$ &
            $R_{25}$      & $I4/mcm$ & $-$1443 \\
          & $\Gamma_{25}$ & $P{\bar 4}m2$ & $-$341 &
            $M_2^\prime$  & $P4/nmm$ & $-$688$^a$ &
	    $M_3$         & $P4/mbm$ & $-$1449 \\
	  & $R_{15}$      & $I4/mmm$ & $-$375$^a$ &
            $X_5^\prime$  & $Pmma$   & $-$752$^a$ &
	    $\Gamma_{25}$ & $R32$    & $-$1486 \\
    	  & $X_5$         & $Pmma$   & $-$447 &
	    $X_5$         & $Cmcm$   & $-$867 &
	    $R_{25}+M_3$  & $Cmcm$   & $-$2036$^a$ \\
          & $X_5^\prime$  & $Cmcm$   & $-$449$^a$ &
            $\Gamma_{15}$ & $R3m$    & $-$868 &
            $R_{15}$, $R_{25}$ & $Imma$   & $-$2215$^a$ \\
          & $R_{15}$      & $R{\bar 3}m$ & $-$465$^a$ &
            $\Gamma_{15}$ & $P4mm$   & $-$1104 &
            $R_{25}$      & $R{\bar 3}c$ & $-$2271 \\
          & $M_5^\prime$  & $Pmma$   & $-$555$^a$ &
            $\Gamma_{15}$, $\Gamma_{25}$ & $Amm2$ & $-$1254 &
            $R_{25}+M_3$  & $Pbnm$   & \textbf{$-$2312} \\
\hline
GeTiO$_3$ & $\Gamma_{25}$ & $P{\bar 4}m2$ & $-$0.5$^a$ &
	    $M_3$         & $P4/mbm$ & $-$444 &
            $R_{25} + M_3$ & $Cmcm$  & $-$733$^a$ \\
	  & $\Gamma_{25}$ & $R32$    & $-$1.5$^a$ &    
            $R_{25}$      & $I4/mcm$ & $-$455 &
	    $R_{25} + M_3$ & $Pbnm$   & $-$810 \\
          & $R_{15}$      & $I4/mmm$ & $-$314$^a$ &
            $R_{15}$      & $R{\bar 3}m$ & $-$501$^a$ &
            $\Gamma_{15}$ & $P4mm$   & $-$854 \\
          & $X_5^\prime$  & $Pmma$   & $-$328 &
	    $R_{25}$      & $R{\bar 3}c$ & $-$589 &
            $\Gamma_{15}$, $\Gamma_{25}$ & $Amm2$ & $-$881$^a$ \\
	  & $X_5^\prime$  & $Cmcm$   & $-$428 &
	    $R_{15}$, $R_{25}$ & $Imma$ & $-$704$^a$ &
	    $\Gamma_{15}$ & $R3m$    & \textbf{$-$1053} \\
\hline
SnTiO$_3$ & $R_{15}$      & $I4/mmm$ & $-$1.1$^a$ &
	    $M_3$         & $P4/mbm$ & $-$57 &
	    $R_{15}$, $R_{25}$ & $Imma$$^b$ & $-$84 \\
	  & $R_{15}$      & $R{\bar 3}m$ & $-$1.7$^a$ &
            $R_{25}$      & $I4/mcm$ & $-$67 &
	    $\Gamma_{15}$ & $R3m$    & $-$240 \\
          & $X_5^\prime$  & $Pmma$   & $-$21 &
	    $R_{25}$      & $R{\bar 3}c$ & $-$74 &
            $\Gamma_{15}$ & $Amm2$   & $-$242$^a$ \\
	  & $X_5^\prime$  & $Cmcm$   & $-$23 &
            $R_{25} + M_3$ & $Cmcm$  & $-$78$^a$ &
            $\Gamma_{15}$ & $P4mm$   & \textbf{$-$291} \\
\hline
PbTiO$_3$ & $M_3$         & $P4/mbm$ & $-$10.1 &
	    $R_{25}$      & $R{\bar 3}c$ & $-$21.6 &
	    $\Gamma_{15}$ & $Amm2$   & $-$70.4$^a$ \\
          & $R_{25}$      & $I4/mcm$ & $-$19.6 &
	    $R_{25}$      & $Imma$$^b$ & $-$22.2 &
            $\Gamma_{15}$ & $P4mm$   & \textbf{$-$84.4} \\
          & $R_{25} + M_3$ & $Cmcm$  & $-$19.7$^a$ &
	    $\Gamma_{15}$ & $R3m$    & $-$66.3 \\
\cline{1-2}
\multicolumn{7}{l}{$^a$\rule{0pt}{11pt}\footnotesize Results of additional
calculations.}\\
\multicolumn{7}{l}{$^b$\rule{0pt}{11pt}\footnotesize The $Pbnm$ structure found
in earlier calculations relaxes to the $Imma$ one.}
\end{tabular}
\end{ruledtabular}
\end{table*}

\begin{table}
\caption{\label{Table4}Frequencies of optical phonons at the $\Gamma$ point
in the cubic phase of \emph{A}TiO$_3$ compounds (in~cm$^{-1}$).}
\begin{ruledtabular}
\begin{tabular}{ccccccccc}
Compound  & Source             & TO1 & TO2 & TO3 & LO1 & LO2 & LO3 & $\Gamma_{25}$ \\
\hline
MgTiO$_3$ & this work          & 260$i$ & 151 & 649 & 106$i$ & 372 & 905 & 191$i$ \\
CaTiO$_3$ & this work          & 165$i$ & 176 & 607 & 122 & 407 & 857 & 93 \\
          & theor.~\cite{PhysRevLett.72.3618}    & 153$i$ & 188 & 610 & 133 & 427 & 866 & --- \\
          & theor.~\cite{PhysRevB.62.3735}    & 140$i$ & 200 & 625 & 136 & 428 & 864 & 130 \\
SrTiO$_3$ & this work          & 68$i$  & 162 & 549 & 152 & 428 & 792 & 202 \\
          & theor.~\cite{PhysRevLett.72.3618}       & 41$i$  & 165 & 546 & 158 & 454 & 829 & --- \\
	  & theor.~\cite{Ferroelectrics.194.109}      & 100$i$ & 151 & 522 & 146 & 439 & 751 & 219 \\
	  & exp.~\cite{PhysRevB.22.5501} & ---  & 175 & 544 & 172 & 475 & 796 & --- \\
BaTiO$_3$ & this work          & 151$i$ & 175 & 471 & 172 & 439 & 683 & 269 \\
          & theor.~\cite{PhysRevLett.72.3618}       & 178$i$ & 177 & 468 & 173 & 453 & 738 & --- \\
          & theor.~\cite{Ferroelectrics.206.205}   & 195$i$ & 166 & 455 & 162 & 434 & 657 & 266 \\
	  & exp.~\cite{Ferroelectrics.137.65}\,$^a$ & ---   & 181 & 487 & 180 & 468 & 717 & 306 \\
RaTiO$_3$ & this work          & 212$i$ & 172 & 444 & 166 & 434 & 638 & 287 \\
\hline
ZnTiO$_3$ & this work          & 240$i$ & 76  & 645 & 105$i$ & 316 & 815 & 353$i$ \\
CdTiO$_3$ & this work          & 187$i$ & 97  & 616 & 34  & 353 & 820 & 231$i$ \\
\hline
GeTiO$_3$ & this work          & 247$i$ & 122 & 583 & 68$i$ & 356 & 762 & 49$i$ \\
SnTiO$_3$ & this work          & 185$i$ & 126 & 505 & 80  & 375 & 689 & 183 \\
PbTiO$_3$ & this work          & 150$i$ & 116 & 499 & 96  & 394 & 693 & 202 \\
          & theor.~\cite{PhysRevLett.72.3618}       & 144$i$ & 121 & 497 & 104 & 410 & 673 & --- \\
	  & theor.~\cite{PhysRevB.55.6161}    & 182$i$ & 63  & 447 & 47  & 418 & 610 & --- \\
\cline{1-2}
\multicolumn{6}{l}{$^a$\rule{0pt}{11pt}\footnotesize
Data for the tetragonal phase.}
\end{tabular}
\end{ruledtabular}
\end{table}

\begin{table}
\caption{\label{Table5}Lowest phonon frequencies at high-symmetry points of
the Brillouin zone in the cubic phase of \emph{A}TiO$_3$ compounds (in~cm$^{-1}$).}
\begin{ruledtabular}
\begin{tabular}{ccccccc}
Compound   & Source      & $\Gamma$ & $X$  & $M$    & $R$    & $\Lambda$ \\
\hline
MgTiO$_3$  & this work   & 260$i$ & 190$i$ & 314$i$ & 315$i$ & 246$i$ \\
CaTiO$_3$  & this work   & 165$i$ & 32$i$  & 215$i$ & 226$i$ & 122$i$ \\
           & theor.~\cite{PhysRevB.62.3735}
	                 & 140$i$ & 20     & 207$i$ & 219$i$ & --- \\
SrTiO$_3$  & this work   & 68$i$  & 98     & 86$i$  & 119$i$ & 100  \\
BaTiO$_3$  & this work   & 151$i$ & 96$i$  & 59$i$  & 134    & 105 \\
           & theor.~\cite{Ferroelectrics.206.205}
                         & 219$i$ & 189$i$ & 167$i$ & 128    & --- \\
RaTiO$_3$  & this work   & 212$i$ & 182$i$ & 158$i$ & 110    & 87  \\
\hline
ZnTiO$_3$  & this work   & 353$i$\,$^a$ & 319$i$ & 437$i$ & 424$i$ & 337$i$ \\
CdTiO$_3$  & this work   & 231$i$\,$^a$ & 184$i$ & 333$i$ & 328$i$ & 265$i$ \\
\hline
GeTiO$_3$  & this work   & 247$i$ & 148$i$ & 254$i$ & 251$i$ & 201$i$ \\
SnTiO$_3$  & this work   & 185$i$ & 56$i$  & 144$i$ & 148$i$ & 97$i$  \\
PbTiO$_3$  & this work   & 150$i$ & 30     & 96$i$  & 113$i$ & 15$i$  \\
           & theor.~\cite{PhysRevB.55.6161}
	                 & 182$i$ & 31$i$  & 35$i$  & 145$i$ & 58$i$ \\
\cline{1-2}
\multicolumn{6}{l}{$^a$\rule{0pt}{11pt}\footnotesize
 The $\Gamma_{25}$ mode.}
\end{tabular}
\end{ruledtabular}
\end{table}

The calculated optical-phonon frequencies are also in good agreement with
available experimental data and calculations performed by other authors
(Tables~\ref{Table4}, \ref{Table5}). The imaginary frequencies in
Tables~\ref{Table4} and \ref{Table5} correspond to unstable modes, the
squares of whose frequencies are negative.

The good agreement of our calculations with the experimental data and the
calculations of other authors for CaTiO$_3$, SrTiO$_3$, BaTiO$_3$, and
PbTiO$_3$ suggests that the proposed approach can be used to predict the
properties of poorly studied and hypothetical%
    \footnote{Our calculations show that the energy of the ilmenite phase of
    SnTiO$_3$, GeTiO$_3$, CdTiO$_3$, ZnTiO$_3$, and MgTiO$_3$ at $T = 0$ is
    lower than the energy of the most stable of the distorted perovskite phases.
    However, the energy difference between these phases is large enough
    (0.30--0.33~eV) only for the two last compounds.}
titanates with the perovskite structure and to discuss the factors causing the
appearance of ferroelectricity in them. The properties of these crystals
calculated for the theoretical lattice parameter (i.e., the value corresponding
to a minimum of the total energy) are given in Tables~\ref{Table3} to \ref{Table7}.
Table~\ref{Table3} presents the energies of different low-symmetry phases
measured relative to the energy of the parent cubic phase.
Table~\ref{Table4} gives the frequencies of optical phonons (three infrared-active
$\Gamma_{15}$ modes and one infrared-inactive $\Gamma_{25}$ mode) in the cubic
phase. Table~\ref{Table5} gives the lowest phonon
frequencies at high-symmetry points of the Brillouin zone. The values of the
Born effective charges and optical dielectric constants for the cubic phase of
\emph{A}TiO$_3$ compounds are given in Table~\ref{Table6}. Finally, the elastic
moduli of several crystals are presented in Table~\ref{Table7}.

\begin{figure*}
\begin{minipage}[c]{.5\textwidth}
\centering
\includegraphics{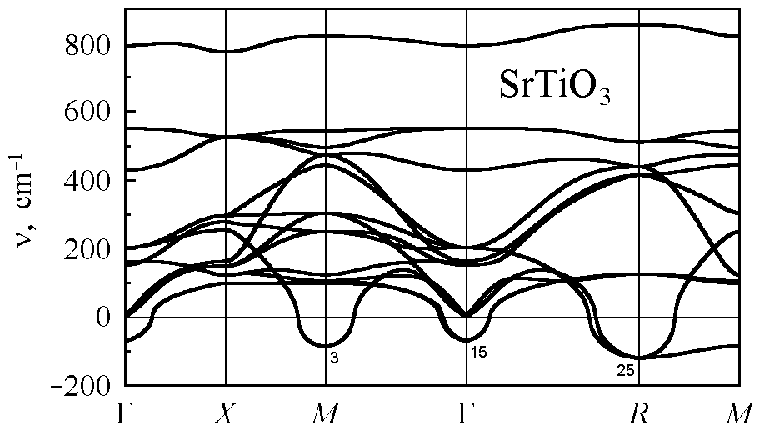}
\end{minipage}%
\begin{minipage}[c]{.5\textwidth}
\centering
\includegraphics{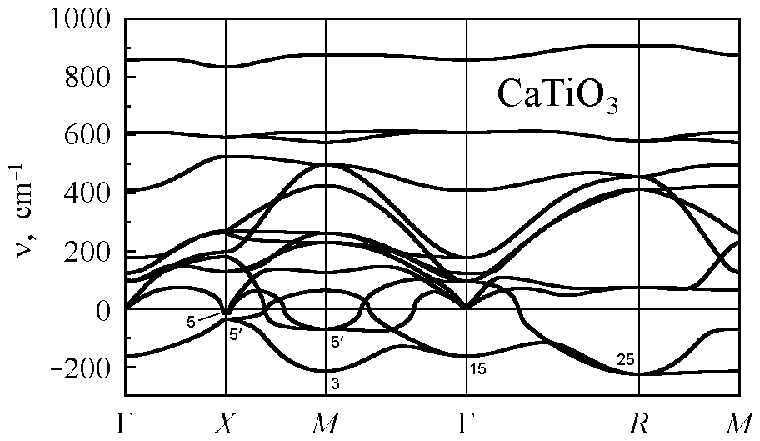}
\end{minipage}

\begin{minipage}[c]{.5\textwidth}
\centering
\includegraphics{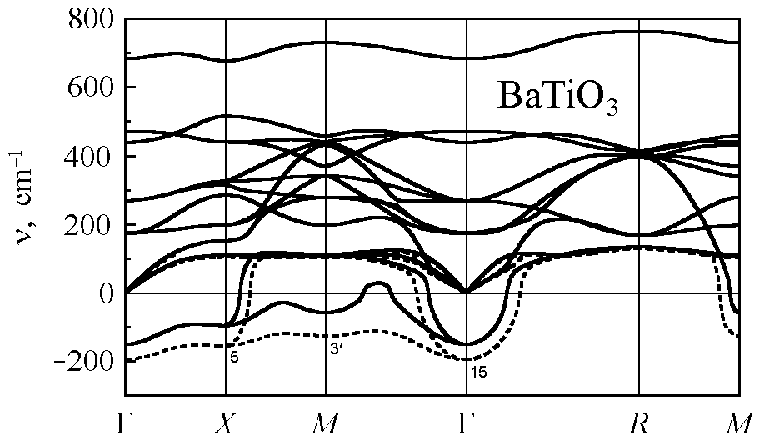}
\end{minipage}%
\begin{minipage}[c]{.5\textwidth}
\centering
\includegraphics{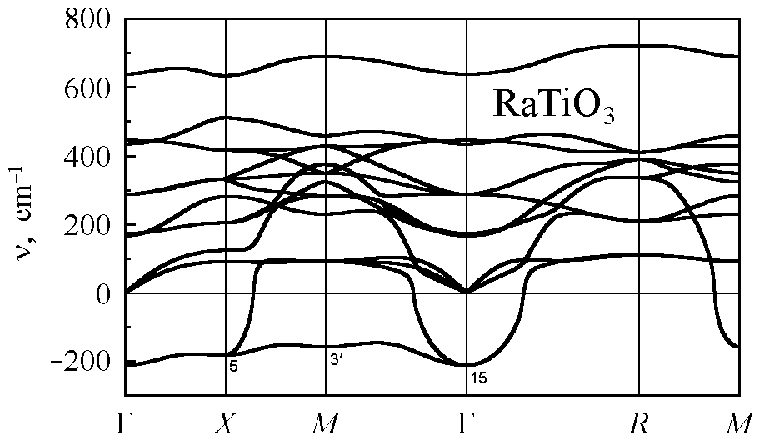}
\end{minipage}

\begin{minipage}[c]{.5\textwidth}
\centering
\includegraphics{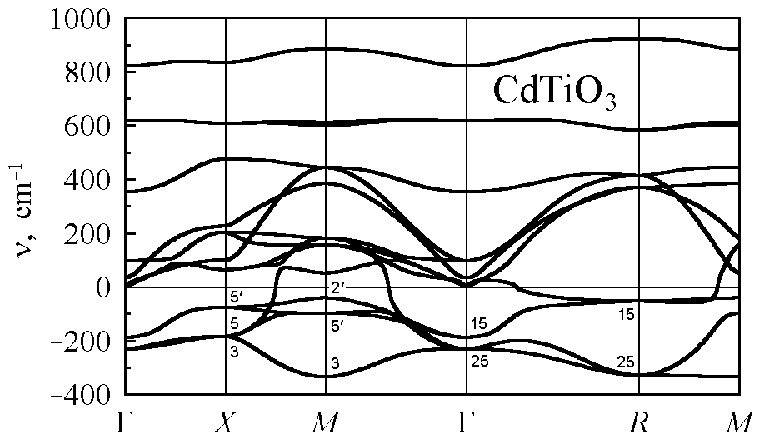}
\end{minipage}%
\begin{minipage}[c]{.5\textwidth}
\centering
\includegraphics{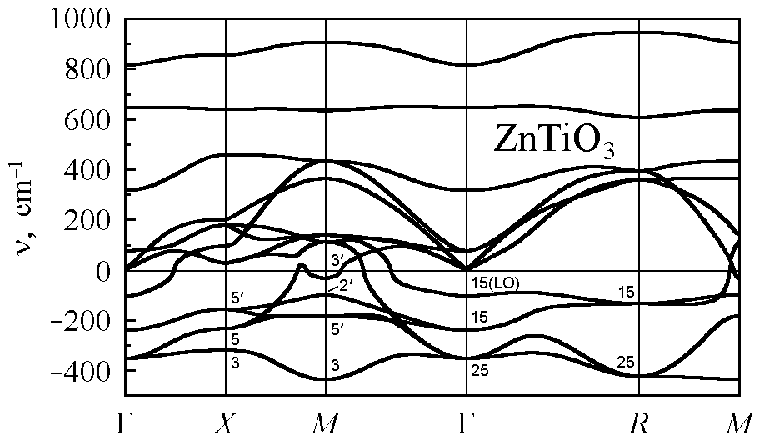}
\end{minipage}

\begin{minipage}[c]{.5\textwidth}
\centering
\includegraphics{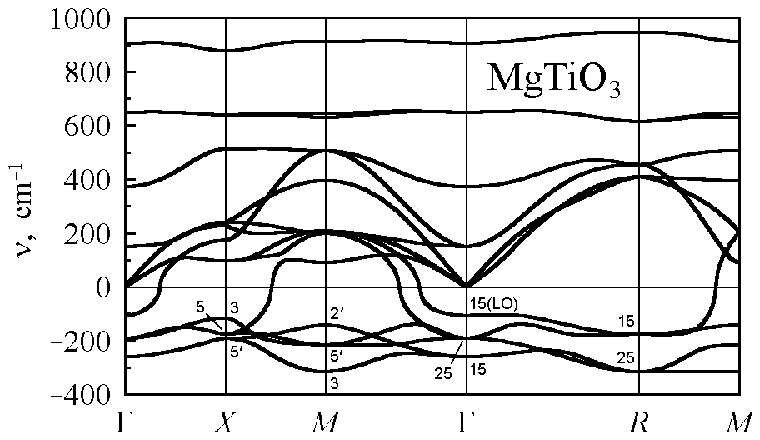}
\end{minipage}%
\begin{minipage}[c]{.5\textwidth}
\centering
\includegraphics{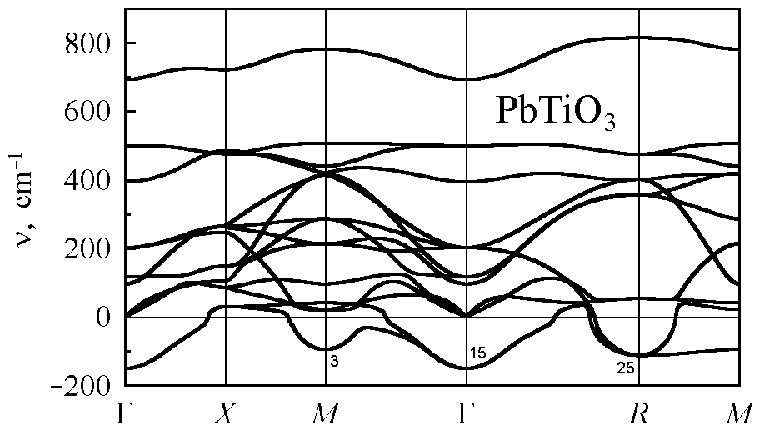}
\end{minipage}

\begin{minipage}[c]{.5\textwidth}
\centering
\includegraphics{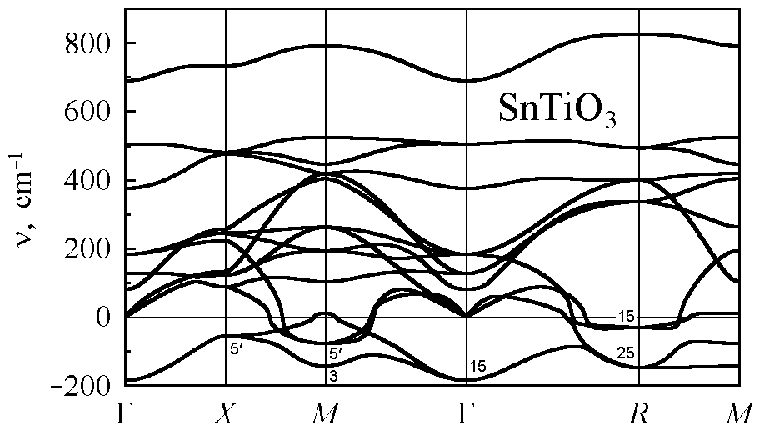}
\end{minipage}%
\begin{minipage}[c]{.5\textwidth}
\centering
\includegraphics{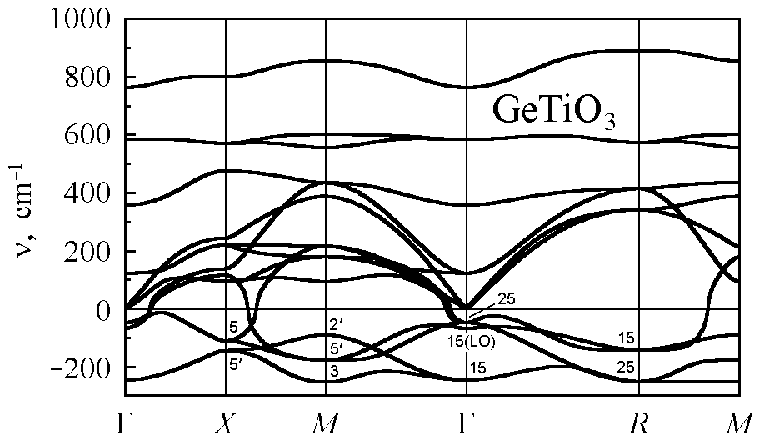}
\end{minipage}
\caption{\label{Fig1}Phonon dispersion curves for the cubic phase of
\emph{A}TiO$_3$ compounds. The labels indicate the symmetry of unstable modes.}
\end{figure*}

The phonon dispersion curves along some high-symmetry directions calculated
for the cubic phase of ten titanates are shown in Fig.~\ref{Fig1}. The imaginary
phonon frequencies associated with the structural instability of the crystals are
represented in Fig.~\ref{Fig1} by negative values.

\begin{table}
\caption{\label{Table6}Effective charges and optical dielectric constant
for the cubic phase of \emph{A}TiO$_3$ compounds.}
\begin{ruledtabular}
\begin{tabular}{cccccc}
Compound  & $Z^*_{\rm A}$ & $Z^*_{\rm Ti}$ & $Z^*_{\rm O\perp}$ & $Z^*_{\rm O\parallel}$ & $\epsilon_\infty$ \\
\hline
MgTiO$_3$ & 2.537 & 7.773 & $-$2.026 & $-$6.258 &  7.01 \\
CaTiO$_3$ & 2.579 & 7.692 & $-$2.085 & $-$6.101 &  6.84 \\
SrTiO$_3$ & 2.561 & 7.725 & $-$2.099 & $-$6.088 &  6.87 \\
BaTiO$_3$ & 2.738 & 7.761 & $-$2.186 & $-$6.128 &  7.28 \\
RaTiO$_3$ & 2.764 & 7.789 & $-$2.181 & $-$6.192 &  7.42 \\
\hline
ZnTiO$_3$ & 3.233 & 8.257 & $-$2.427 & $-$6.637 & 11.64 \\
CdTiO$_3$ & 3.040 & 8.052 & $-$2.300 & $-$6.493 &  9.35 \\
\hline
GeTiO$_3$ & 4.460 & 7.572 & $-$2.860 & $-$6.314 & 10.49 \\
SnTiO$_3$ & 4.255 & 7.529 & $-$2.745 & $-$6.294 & 10.18 \\
PbTiO$_3$ & 3.931 & 7.623 & $-$2.635 & $-$6.283 &  9.34 \\
\end{tabular}
\end{ruledtabular}
\end{table}

\begin{table}
\caption{\label{Table7}Elastic moduli of the cubic phase of
\emph{A}TiO$_3$ compounds (in GPa).}
\begin{ruledtabular}
\begin{tabular}{cccccc}
Compound  & Source      & $C_{11}$ & $C_{12}$ & $C_{44}$ & $B$ \\
\hline
CaTiO$_3$ & this work   & 388     & 100     & 91      & 196 \\
          & theor.~\cite{PhysRevB.49.5828} & 407 & 96     & 101     & 200 \\
SrTiO$_3$ & this work   & 373     & 103     & 108     & 193 \\
          & theor.~\cite{PhysRevB.49.5828} & 388 & 104    & 117     & 199 \\
          & exp.~\cite{LB}  & 316--348     & 101--103 & 119--124 & 174--183 \\
BaTiO$_3$ & this work   & 338     & 110     & 123     & 186 \\
          & theor.~\cite{PhysRevB.49.5828} & 329 & 117    & 130     & 188 \\
%          & exp.~?     &         &         &         & (139--195) \\
RaTiO$_3$ & this work   & 319     & 112     & 126     & 181 \\
PbTiO$_3$ & this work   & 336     & 127     & 95      & 197 \\
          & theor.~\cite{PhysRevB.49.5828} & 334 & 145 & 100   & 208 \\
          & theor.~\cite{PhysRevB.55.6161} & 320 & 141 & 375?  & 203 \\
\end{tabular}
\end{ruledtabular}
\end{table}

\section{Results}

As follows from Fig.~\ref{Fig1}, unstable optical modes of different symmetry
are present in the phonon spectra of all studied titanates. At first, we discuss
the phonon spectra of well-studied compounds.

A specific feature of the dispersion curves of SrTiO$_3$ is that among three
unstable phonons at the $\Gamma$, $R$, and $M$ points, the most unstable phonon
is that at the $R$ point (the $R_{25}$ mode)%
    \footnote{We use the mode notation introduced in Ref.~\onlinecite{PhysRev.134.A981}.}
and that the phonon frequency depends only weakly on the wave vector along
the $R$--$M$ line. As shown in Ref.~\onlinecite{Ferroelectrics.194.109}, in the
real space, these unstable phonons with the wave vectors near the edges of the
cubic Brillouin zone correspond to the rotation of the oxygen octahedra linked
together by shared vertices, with the correlation length for rotations being
three to five lattice periods. Thus, the $R_{25}$ mode and the $M_3$ mode (which
corresponds to the unstable phonon at the $M$ point) describe the structural
distortions associated with the rotation of the octahedra. The unstable
$\Gamma_{15}$ phonon mode at the $\Gamma$ point corresponds to the ferroelectric
distortion of the structure.

The $R_{25}$ mode is triply degenerate, and the distortions described by the
three-component order parameters ($\eta$, 0, 0), ($\eta$, $\eta$, 0), and
($\eta$, $\eta$, $\eta$) lead to low-symmetry phases with the space groups
$I4/mcm$, $Imma$, and $R{\bar 3}c$, respectively. The $M_3$ mode is nondegenerate,
and its condensation lowers the crystal symmetry to $P4/mbm$. The triply
degenerate $\Gamma_{15}$ mode results in the well-known space groups $P4mm$, $Amm2$,
and $R3m$. From comparing the energies of these phases (Table~\ref{Table3}),
%    \footnote{We do not present in the table the energies of phases described
%    by the order parameter ($\eta$, $\eta$, 0) because these phases occur
%    in a rare case when the coefficient of the second fourth-order invariant
%    constructed from the order parameter components becomes zero.}
it follows that the lowest-energy phase in SrTiO$_3$ is the $I4/mcm$ one, into
which strontium titanate transforms with decreasing temperature. The instability
of the ferroelectric mode in this compound is not sufficiently strong for
ferroelectricity to occur.

In CaTiO$_3$, in addition to the instabilities found above, three weak
antiferroelectric-type instabilities associated with the $X_5$, $X'_5$, and
$M'_5$ modes arise and the $R$--$M$ segment of the phonon spectrum becomes
practically dispersionless (compare mode energies in Table~\ref{Table5}). In the
latter case, as shown by theoretical calculations,~\cite{JChemPhys.114.2395}
the simultaneous condensation of two unstable $R_{25}$ and $M_3$ modes results
in the formation of the $Pbnm$ phase having the lowest energy
among the possible distorted phases (Table~\ref{Table3}).%
    \footnote{In the case of SrTiO$_3$, despite the presence of the unstable
    $R_{25}$ and $M_3$ modes in the phonon spectrum of the cubic phase, the
    $M_3$ instability disappears when switching on the $R_{25}$ rotations, and
    the energy of the $Pbnm$ phase (which relaxes to the $Imma$ phase) is 2~meV
    higher than that of the $I4/mcm$ phase.}
The transition from the high-temperature $Pm3m$ phase to the low-temperature $Pbnm$ phase can
occur through one of three intermediate phases ($P4/mbm$, $I4/mcm$, and $R{\bar 3}c$),
whose energies are 0.11--0.17~eV higher than the energy of the $Pbnm$ phase.%
    \footnote{Further calculations have revealed two additional phases, $Cmcm$
    and $Imma$, which are closer in energy to the ground state (Table~\ref{Table5}).}
The ferroelectric $P4mm$ and $R3m$ phases in CaTiO$_3$ have much higher energies
and never arise. As for the weakly unstable $X'_5$, $X_5$, and $M'_5$ modes,
they are doubly degenerate and the distortions described by the order parameters
($\eta$, 0) and ($\eta$, $\eta$) result in the $Pmma$, $Cmcm$, and $Cmmm$ phases.
The energy gain resulting from the transformation into these phases, however,
does not exceed 7~meV.

The phonon spectrum of BaTiO$_3$ significantly differs from the spectra
discussed above by the absence of instability at the $R$ point and the
appearance of highly unstable $X_5$ and $M'_3$ modes (at the $X$ and $M$
points, respectively) corresponding to the antiferroelectric distortions of
the structure into the $Pmma$, $Cmcm$, and $P4/nmm$ phases. Among the unstable
modes, the ferroelectric $\Gamma_{15}$ mode is the most unstable and it
determines the distortions observed in the crystal (the energies of three
above-mentioned antiferroelectric phases are higher than that of the polar
$P4mm$ phase). We note that the weaker phonon instability in our calculations
as compared to the results of Ref.~\onlinecite{PhysRevB.60.836} is because
our calculations were performed for the \emph{theoretical} lattice
parameter, whereas the calculations of Ref.~\onlinecite{PhysRevB.60.836}
were performed for its \emph{experimental} value. To illustrate the
influence of the lattice strain on the phonon spectrum, the low-energy
part of this spectrum calculated for the same lattice parameter as in
Ref.~\onlinecite{PhysRevB.60.836} is shown in Fig.~\ref{Fig1} by the
dashed line.

The weak dependence of the unstable TO phonon frequency on the wave
vector along the $\Gamma$--$X$--$M$--$\Gamma$ path for vibrations polarized
along the fourfold axes of the cubic structure was first discovered in
Ref.~\onlinecite{PhysRevLett.74.4067}. This dependence gives evidence for
the dominant role of the vibrations in the \mbox{...--O--Ti--O--...} linear
chains oriented along these axes and for the weak interaction between
adjacent chains.

A comparison of the phonon spectra of BaTiO$_3$ and RaTiO$_3$ shows that these
spectra are very similar. In radium titanate, the most unstable mode is the
$\Gamma_{15}$ mode whose frequency depends only weakly on the wave vector along
the $\Gamma$--$X$--$M$-$\Gamma$ path. The instability of phonons in RaTiO$_3$
is even more pronounced than in BaTiO$_3$. Taking into account the calculated
energies of the distorted phases (Table~\ref{Table3}), we can suppose that
RaTiO$_3$ is also a ferroelectric and that, as the temperature decreases, it
undergoes three successive phase transitions as barium titanate does. The
temperatures of these transitions are likely to exceed those in barium titanate.
The values of the spontaneous polarization calculated by the Berry phase method
in RaTiO$_3$ are also higher than those in BaTiO$_3$; they are 0.36~C/m$^2$
in the tetragonal phase and 0.41~C/m$^2$ in the rhombohedral phase. The
calculated elastic moduli of the cubic RaTiO$_3$ are given in Table~\ref{Table7}.

Now, we consider the phonon spectrum of CdTiO$_3$. This spectrum has a number
of unstable modes at the $X$, $M$, and $R$ points ($X_3$, $X_5$, $X'_5$, $M_3$,
$M'_5$, $M'_2$, $R_{25}$, $R_{15}$ modes) and two unstable modes at the $\Gamma$
point. It is surprising that the main instability at the $\Gamma$ point is due
to the $\Gamma_{25}$ mode associated with the deformation of the oxygen
octahedron rather than to the ferroelectric $\Gamma_{15}$ mode (see mode
frequencies in Table~\ref{Table4}). This deformation can result in the
formation of the $P{\bar 4}m2$, $Amm2$, and $R32$ phases, depending on the
number of nonzero components of the order parameter.%
    \footnote{The fact that the lattice symmetry is lowered to the polar
    $Amm2$ group follows from the transformation properties of the order
    parameter ($\eta$, $\eta$, 0). The spontaneous polarization in this
    phase is 0.018 C/m$^2$.}
The energy of the most stable of these phases ($R32$, Table~\ref{Table3}) is
lower than the energy of the polar phases. Because of the qualitative similarity
between the phonon spectra of calcium and cadmium titanates and between
the eigenvectors of their unstable modes at the $R$ and $M$ points, CdTiO$_3$
can be considered as an analog of CaTiO$_3$ characterized by an even higher
instability. Therefore, at room temperature, the structure of its nonpolar
phase is $Pbnm$, as in the case of CaTiO$_3$. This conclusion was also made
in Ref.~\onlinecite{PhysRevB.66.233106}. According to our data, the energy of
this phase is lower than that of the cubic phase by 1.28~eV (Table~\ref{Table3}),
which is somewhat higher than the value obtained in
Ref.~\onlinecite{PhysRevB.66.233106} (0.8~eV) and
Ref.~\onlinecite{ApplPhysLett.81.3443} (0.91~eV).

Although the ferroelectric instability associated with the $\Gamma_{15}$ mode
is not very important in the cubic CdTiO$_3$, it is known that this instability
exists in the $Pbnm$ phase and results in the ferroelectric phase transition at
80~K. The first-principles calculations of the properties of the orthorhombic
$Pbnm$ phase~\cite{PhysRevB.66.233106} did not found a stable ferroelectric
distortion in it. In contrast, our calculations of the phonon spectrum at the
$\Gamma$ point of the orthorhombic CdTiO$_3$ reveal two unstable $B_{1u}$ and
$B_{2u}$ modes resulting in the formation of polar $Pb2_1m$ and $Pbn2_1$ phases,
respectively. These lattice distortions have been observed in X-ray studies of
cadmium titanate at low temperatures.~\cite{Ferroelectrics.259.85,Ferroelectrics.284.107}
The properties of these phases will be considered in a separate paper.%
    \footnote{This paper has been published in Phys. Solid State \textbf{51},
    802 (2009); \texttt{DOI: 10.1134/S1063783409040283}. It has been shown that
    only the $Pbn2_1$ phase is stable at low temperatures; the other phase is
    suppressed by zero-point vibrations.}

The phonon spectrum of ZnTiO$_3$ is qualitatively similar to that of CdTiO$_3$,
but it has an additional weak unstable $M'_3$ mode and is even less stable. The
$\Gamma_{25}$ mode in it is also less stable than the ferroelectric
$\Gamma_{15}$ mode (Table~\ref{Table4}). However, as the most unstable modes
in ZnTiO$_3$ are the $R_{25}$ and $M_3$ ones, the $Pbnm$ phase has the lowest
energy (Table~\ref{Table3}). The calculations of the phonon spectrum at
the $\Gamma$ point of the orthorhombic zinc titanate reveal two unstable
$B_{1u}$ and $B_{2u}$ modes, which can cause the formation of the same polar
phases as in cadmium titanate.

The phonon spectrum of MgTiO$_3$ is intermediate between those of zinc titanate
and calcium titanate. It also has unstable $\Gamma_{15}$ and $\Gamma_{25}$
modes, but in magnesium titanate the ferroelectric $\Gamma_{15}$ mode is less
stable (Table~\ref{Table4}). Nevertheless, as the phonons at the $R$ and $M$
points have the lowest frequency, the $Pbnm$ phase is the most energetically
favorable (Table~\ref{Table3}). The calculations of the phonon spectrum at
the $\Gamma$ point in the orthorhombic magnesium titanate reveal one unstable
$B_{1u}$ mode, which can cause the $Pbnm \to Pbn2_1$ ferroelectric phase transition.

Finally, we discuss the phonon spectra of PbTiO$_3$, SnTiO$_3$, and GeTiO$_3$.
The ferroelectric instability in these three compounds is associated with the
$\Gamma_{15}$ mode which competes with the unstable $R_{25}$ and $M_3$ modes.
From comparing the energies of different distorted phases (Table~\ref{Table3}),
it follows that even in GeTiO$_3$, in which the unstable phonons at the $\Gamma$,
$R$, and $M$ points are close in frequency, the ferroelectric instability is
dominant. The calculated spontaneous polarization is 1.28~C/m$^2$ in the
tetragonal SnTiO$_3$ and 1.37~C/m$^2$ in the rhombohedral GeTiO$_3$. Our value
of the spontaneous polarization in SnTiO$_3$ is significantly higher than the
value of 0.73~C/m$^2$ obtained in Ref.~\onlinecite{MRSProc.748.U3.13}. Among the
perovskite compounds studied to date, GeTiO$_3$ is likely to have the highest
spontaneous polarization.

It should be noted that in PbTiO$_3$ and SnTiO$_3$, the most energetically
favorable phase is the tetragonal $P4mm$ phase, whereas in GeTiO$_3$ it is
the rhombohedral $R3m$ phase. At the same time, the lattice strain in the
tetragonal GeTiO$_3$ ($c/a = 1.1821$) is much higher than in lead titanate
($c/a = 1.0590$). This result calls into question the
conclusion~\cite{Nature.358.136} that the stabilization of the tetragonal phase
in PbTiO$_3$ is due to a high lattice strain (large $c/a$ ratio).

\section{Discussion}

As follows from Fig.~\ref{Fig1}, the phonon spectra of all \emph{A}TiO$_3$
perovskite crystals studied in this work are characterized by several unstable
modes, one of which is always the ferroelectric $\Gamma_{15}$ mode. In the
case where the $R_{25}$ and $M_3$ modes competing with it have a lower
frequency, the crystal undergoes distortions---the octahedra rotations,
and its symmetry is lowered to $I4/mcm$ or $Pbnm$. The tendency toward such
structural phase transitions increases with decreasing the $A$ atom size.

From analyzing the characteristics of the $\Gamma_{15}$ mode, one can draw a
conclusion on the nature of the ferroelectric instability in studied crystals.
As mentioned above, the dispersion law of this mode in BaTiO$_3$ and RaTiO$_3$
indicates strongly correlated motion of atoms along the ...--O--Ti--O--...
chains. The analysis of the $\Gamma_{15}$ phonon eigenvectors (Table~\ref{Table8})
shows that the $A$ atoms do not contribute much to the motion which
is mainly determined by the out-of-phase Ti and O$_\parallel$ displacements. As
the size of the $A$ atoms decreases, their contribution to the motion increases
and becomes dominant, whereas the contribution of the Ti atoms decreases and
the out-of-phase motion involves now not O$_{\parallel}$ but O$_{\perp}$
displacements. Thus, in crystals with small $A$ atoms, the ferroelectric mode
is determined by the out-of-phase displacements of the $A$ atom and the
cuboctahedron of oxygen atoms.

\begin{table}
\caption{\label{Table8}Eigenvectors of the dynamic matrix for an unstable TO1
phonon at the $\Gamma$ point in the cubic phase of \emph{A}TiO$_3$ compounds.}
\begin{ruledtabular}
\begin{tabular}{ccccc}
Compound  & $x_{\rm A}$ & $x_{\rm Ti}$ & $x_{\rm O\perp}$ & $x_{\rm O\parallel}$ \\
\hline
MgTiO$_3$ & +0.6828 & +0.1831 & $-$0.4800 & $-$0.1985 \\
CaTiO$_3$ & +0.5693 & +0.2391 & $-$0.5225 & $-$0.2696 \\
SrTiO$_3$ & +0.3434 & +0.3852 & $-$0.5372 & $-$0.3956 \\
BaTiO$_3$ & +0.0299 & +0.6734 & $-$0.3561 & $-$0.5404 \\
RaTiO$_3$ & +0.0051 & +0.6750 & $-$0.2841 & $-$0.6188 \\
\hline
ZnTiO$_3$ & +0.5167 & +0.1889 & $-$0.5655 & $-$0.2403 \\
CdTiO$_3$ & +0.4012 & +0.2358 & $-$0.5919 & $-$0.2875 \\
\hline
GeTiO$_3$ & +0.5367 & +0.1382 & $-$0.5573 & $-$0.2677 \\
SnTiO$_3$ & +0.4177 & +0.2123 & $-$0.5670 & $-$0.3709 \\
PbTiO$_3$ & +0.2973 & +0.2865 & $-$0.5675 & $-$0.4305 \\
\end{tabular}
\end{ruledtabular}
\end{table}

\begin{table}
\caption{\label{Table9} Diagonal elements of the on-site force-constant matrix
$\Phi_{xx}(0, 0)$ for $A$ and Ti atoms in the cubic phase of
\emph{A}TiO$_3$ compounds (in Ha/Bohr$^2$).}
\begin{ruledtabular}
\begin{tabular}{ccc}
Compound  & $A$ atom  & Ti atom \\
\hline
MgTiO$_3$ & $-$0.0109 & +0.1431 \\
CaTiO$_3$ &   +0.0163 & +0.1370 \\
SrTiO$_3$ &   +0.0445 & +0.1196 \\
BaTiO$_3$ &   +0.0755 & +0.0873 \\
RaTiO$_3$ &   +0.0856 & +0.0750 \\
\hline
ZnTiO$_3$ & $-$0.0229 & +0.1072 \\
CdTiO$_3$ & $-$0.0008 & +0.1113 \\
\hline
GeTiO$_3$ & $-$0.0150 & +0.0949 \\
SnTiO$_3$ &   +0.0132 & +0.0786 \\
PbTiO$_3$ &   +0.0269 & +0.0803 \\
\end{tabular}
\end{ruledtabular}
\end{table}

The values of the diagonal elements of the on-site force-constant matrix
$\Phi_{xx}(0, 0)$ for $A$ and Ti atoms are presented in Table~\ref{Table9}.
These matrices are defined in terms of the restoring force acting on an atom
displaced from its site, with the other atoms fixed at their lattice sites.
To determine the on-site force constants from the force constants
calculated using the ABINIT software for a sublattice displaced as a
whole, the force constants calculated on a regular mesh of wave vectors were
averaged.~\cite{PhysRevB.43.7231,PhysRevB.55.10355,PhysRevB.50.13035}
Positive values of the on-site force constants indicate that the position of
an atom at its site is stable, while the negative values indicate that the atom
is off-center. As follows from Table~\ref{Table9}, the off-centering of the $A$
atoms should be observed in \emph{A}TiO$_3$ perovskites for \emph{A} = Mg, Zn,
Cd, and Ge. The Sn, Ca, and Pb atoms
are fairly close to the stability limit against the transition to an
off-center position.

It should be recalled that in this work, the calculations were performed for
the theoretical lattice parameter (corresponding to a minimum of the total
energy). As the systematically underestimated lattice parameters in the LDA
weakens the ferroelectric instability, many authors perform calculations for
the experimental lattice parameters. In order to estimate the influence of this
systematic error, we carried out a computer simulation which showed that an
increase in the lattice parameter by 1\%
(which is a typical LDA error) decreases $\Phi_{xx}(0, 0)$ in PbTiO$_3$
by 0.006~Ha/Bohr$^2$ for the $A$ atom and by 0.016~Ha/Bohr$^2$ for the Ti atom.
As a result, the atoms with $\Phi_{xx}(0, 0)$ close to the stability limit
against the
transition to an off-center position can actually be off-center. Perhaps,
this occurs in lead titanate, as indicated by extended X-ray-absorption
fine-structure (EXAFS) studies.~\cite{PhysRevB.50.13168}

\begin{figure}
\centering
\includegraphics[scale=0.92]{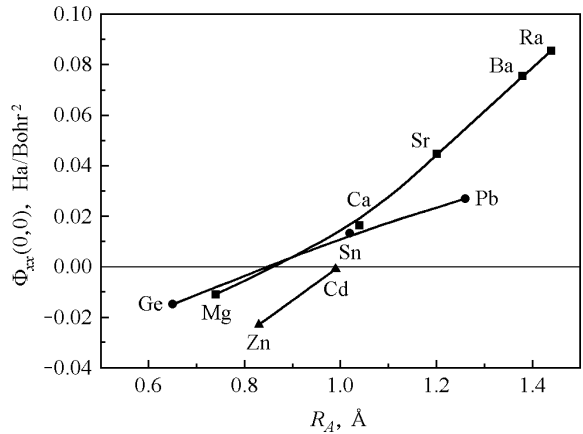}
\caption{\label{Fig2}Dependence of the $\Phi_{xx}(0,0)$ matrix element for
the \emph{A} atom on its ionic radius.}
\end{figure}

The main parameter determining the tendency of the $A$ atom in
\emph{A}TiO$_3$ compounds to become off-center is its atomic size. However, if
one plots the dependence of $\Phi_{xx}(0, 0)$ on the ionic radius of the
\emph{A} atom (Fig.~\ref{Fig2}), it becomes clear that the curve for Zn and
Cd atoms, as well as the curve for Ge, Sn, and Pb atoms, are different from
that for the ``main'' Mg--Ca--Sr--Ba--Ra series.
This difference is likely a consequence of different configuration of the
outer electron shell for the atoms. This configuration is $d^{10}$ for Zn and
Cd; $d^{10}s^2$ for Ge, Sn, and Pb; and $s^2p^6$ for the atoms of the main
series. The difference in the properties of these groups of atoms is clearly
manifested in the Born effective charges of the $A$ atoms
(Table~\ref{Table6}). Indeed, for the main series, the effective charge $Z^*_A$
differs only slightly from the nominal charge of the cation (which indicates
the predominantly ionic character of the $A$--O bond); for the other two groups,
the charge $Z^*_A$ is significantly larger, which indicates that the bonding
becomes more covalent in character.~\cite{PhysRevLett.72.3618}

The obtained results suggest that off-center impurity atoms can exist in solid
solutions of titanates with the perovskite structure. Since the average
interatomic distance in these crystals is determined by the matrix, one can
expect that the atoms for
which $\Phi_{xx}(0, 0)$ has negative or small positive values can be
off-center. Therefore, it is possible that the ferroelectric phase transition
induced by Ca, Cd, and Pb impurities in SrTiO$_3$~(Ref.~\onlinecite{Lemanov2000})
is due to off-centering of these atoms. According to the EXAFS data,
the Ba atoms in SrTiO$_3$ are on-center,~\cite{PhysRevB.62.2969} but the Pb
atoms in SrTiO$_3$ and BaTiO$_3$ can be off-center.~\cite{PhysSolidState.51.991}

The results of this study differ somewhat from those obtained by
Kvyatkovskii,~\cite{PhysSolidState.44.1135} according to which the multiwell
adiabatic potential occurred only for Mg and Zn atoms, whereas Cd atoms remained
in the on-site positions. The discrepancy between our results is likely due to
the fact that in Ref.~\onlinecite{PhysSolidState.44.1135} the calculations were
performed for relatively small clusters in which the extended correlation of
atomic motion cannot be correctly taken into account (according to
Refs.~\onlinecite{PhysRevLett.74.4067} and \onlinecite{Ferroelectrics.206.205},
the correlation length can be as large as 20~{\AA}).

\section{Conclusions}

The pseudopotentials constructed in this work have been used to calculate the
phonon spectra of \emph{A}TiO$_3$ perovskite crystals within the density
functional theory. All known results on the structural instability in these
compounds were reproduced and the properties of new, previously unknown systems
were predicted. The analysis of the phonon spectra, the force constant matrix,
and the eigenvectors of unstable TO phonons enabled to establish the relative
contributions of the chain instability and off-centering of atoms to the
appearance of ferroelectricity in these compounds. The main factors determining
the possible off-centering of the $A$ atoms are the geometric size of these
atoms and the configuration of their outer electron shell.

% Create the reference section using BibTeX:
%\bibliography{all.bib}
%merlin.mbs apsrev4-1.bst 2010-07-25 4.21a (PWD, AO, DPC) hacked
%Control: key (0)
%Control: author (72) initials jnrlst
%Control: editor formatted (1) identically to author
%Control: production of article title (-1) disabled
%Control: page (0) single
%Control: year (1) truncated
%Control: production of eprint (0) enabled
\providecommand{\BIBYu}{Yu}

\end{document}